\documentclass[aps, prb, onecolumn, showkeys, notitlepage]{revtex4-1}

\usepackage{chemformula} 
\usepackage{siunitx} 
\usepackage[inline,shortlabels]{enumitem} 
\usepackage{hyperref} 
\usepackage{graphicx} 
\usepackage{tabularx}
\usepackage{threeparttablex} 
\usepackage[section]{placeins}
\usepackage{multirow}
\usepackage{fancyhdr}
\usepackage{dblfloatfix}
\usepackage{float}

\hypersetup{hidelinks}

\usepackage{natbib}

\makeatletter
\renewcommand{\fnum@figure}{Figure \thefigure}
\makeatother

\usepackage[justification=raggedright, labelsep=period]{caption}

\begin{document}

    \title{ High-Performance Thermoelectric Oxides Based on Spinel Structure}
    \author{M. Hussein N. Assadi}
    \email{h.assadi.2008@ieee.org}
    \affiliation{School of Materials Science and Engineering, UNSW Sydney, NSW, 2052, Australia.}
    \author{J. Julio Guti\'errez Moreno}
    \affiliation{Institute for Advanced Study, Shenzhen University, Shenzhen 518060, China,}
    \affiliation{Key Laboratory of Optoelectronic Devices and Systems of Ministry of Education and Guangdong Province, College of Physics and Optoelectronic Engineering, Shenzhen University, Shenzhen 518060, China.}
     \author{Marco Fronzi}
    \affiliation{School of Mathematical and Physical Science, University of Technology Sydney, Sydney, NSW 2007, Australia.}
    \date{2020}
    
    \begin{abstract}
    High-performance thermoelectric oxides could offer a great energy solution for integrated and embedded applications in sensing and electronics industries. Oxides, however, often suffer from low Seebeck coefficient when compared with other classes of thermoelectric materials. In search of high-performance thermoelectric oxides, we present a comprehensive density functional investigation, based on GGA$+U$ formalism, surveying the 3d and 4d transition-metal-containing ferrites of the spinel structure. Consequently, we predict \ch{MnFe2O4} and \ch{RhFe2O4} have Seebeck coefficients of $\sim \pm 600$ \si{\micro\volt\per\kelvin} at near room temperature, achieved by light hole and electron doping. Furthermore, \ch{CrFe2O4} and \ch{MoFe2O4} have even higher ambient Seebeck coefficients at $\sim \pm 700$ \si{\micro\volt\per\kelvin}. In the latter compounds, the Seebeck coefficient is approximately a flat function of temperature up to $\sim700$ \si{\kelvin}, offering a tremendous operational convenience. Additionally, \ch{MoFe2O4} doped with $10^{19}$ holes $\mathbin{/}$\si{\cm\cubed} has a calculated thermoelectric power factor of $689.81$ \si{\micro\W\per\K\squared\per\m} at 300 \si{\K}, and $455.67$ \si{\micro\W\per\K\squared\per\m} at $600$ \si{\K}. The thermoelectric properties predicted here can bring these thermoelectric oxides to applications at lower temperatures traditionally fulfilled by more toxic and otherwise burdensome materials.
    \end{abstract}
    \keywords{thermoelectric oxides, \ch{MoFe2O4}, \ch{CrFe2O4}, ferrites, spinels, high Seebeck coefficient, density functional theory, Boltzmann transport equation.}

\maketitle

       \section{INTRODUCTION}
Thermoelectric (TE) materials \cite{Snyder2008, Fergus2012, Liu2018} have the potential to fulfill a grand promise for a variety of applications from recovering waste heat in industrial processes \cite{LeBlanc2014, Fachini2019} to powering small autonomous sensors and devices.\cite{Zhang2018} Currently, a wide range of materials including complex chalcogenides\cite{Zhang2017} (compounds containing group VIA elements), skutterudites \cite{Snyder2016} (As-based compounds), half-Heusler alloys \cite{Zhu2015, Zeier2016, Huang2016} (ternary cubic metallic alloys), silicon-germanium based compounds\cite{Nozariasbmarz2017, Lan2017} are considered to be the best performing TE materials. Each class of these TE materials, however, suffers from some shortcomings. Examples include the instability and Se loss throughout the heating/cooling cycles for chalcogenides \ch{CuSe2}\cite{Bohra2016} and \ch{SnSe},\cite{Shi2018} the low or asymmetric dopability in \ch{ZnSb}\cite{Bjerg2012, Niedziolka2014} and \ch{Mg2Si}\cite{Tani2005}, and the criticality and toxicity of \ch{Te} and \ch{Pb} in \ch{PbTe},\cite{Amatya2012} to mention few common TE compounds. One plausible solution to circumvent most of these problems is developing oxide thermoelectric materials. Oxides, having dominantly ionic characters, are chemically more suitable than other thermoelectric materials by two means: (a) a wide range of elements can be doped into these materials; (b) they have higher chemical stability in oxidizing environments. Furthermore, the top-performing oxide thermoelectric materials enjoy the potential of seamless integration with the current oxide electronics\cite{Lorenz2016} for embedded applications,\cite{Haras2018} an advantage not shared with other class of TE materials that require fundamentally different synthesis techniques.

In thermoelectric materials, the Seebeck effect\cite{Orr2016} refers to an electric potential difference ($\Delta V$) created by a temperature gradient ($\Delta T$) across the length of the material itself and quantified by the Seebeck coefficient $S = - \Delta V \mathbin{/} \Delta T$ which is commonly measured in \si{\micro\V\per\K}. $S$ is related to the TE figure of merit ($ZT$), which determines the thermoelectric efficiency of a material by$ ZT = S^2 \sigma T \mathbin{/}\kappa$, where $\sigma$ is the electrical conductivity, $T$ is the absolute temperature, and $\kappa$ is the thermal conductivity of which the electronic contribution is denoted $\kappa_{\mathrm{e}}$. One of the easiest ways to maximize the thermoelectric $ZT$ is first to identify materials with high $S$, and subsequently enhance $S$ through elemental doping. Optimizing $ZT$ is a challenge in itself due to the interdependence of $S$, $\sigma$, and $\kappa$. Many industrial applications also require the thermoelectric materials to maintain their $ZT$ during the operation at varying temperature ranges and when under stress.\cite{Feng2018, Champier2017} Consequently, many factors must be carefully taken into account and fine-tuned in designing new thermoelectric oxides. Given the complexity of such a design, an experimental approach based on judicious guesswork followed by trial and error is cost-prohibitive.

Computational screening has recently emerged as a novel tool in the discovery of brand-new thermoelectric materials.\cite{Curtarolo2013, Assadi2015, Gorai2016, Assadi2017, Gorai2017, Mukherjee2020} Subsequently, through computational survey into uncharted materials’ territory, one may find a desirable oxide alternative to the common thermoelectric materials, at least for those applications where exposure to elements is inevitable. More specifically, our study was motivated by the recent prediction,\cite{Bouhemadou2019} and observation\cite{Maki2016} of spinel oxides with high Seebeck coefficient, and high-performance thermoelectric cubic oxides.\cite{Azough2019} Therefore, in this work, we surveyed a specific class of spinel ferrites, isomorphic to magnetite, in search of high $S$. In particular, our survey spanned twelve 3d and 4d transition metal (TM) containing spinel ferrites, where the TM ions are tetrahedrally coordinated by O, A site, while the Fe ions are octahedrally coordinated, B site (Figure \ref{figure:1}).
       \section{COMPUTATIONAL SETTINGS}
\begin{figure*}[!hbt]
            \centering
            \includegraphics[width=1\columnwidth]{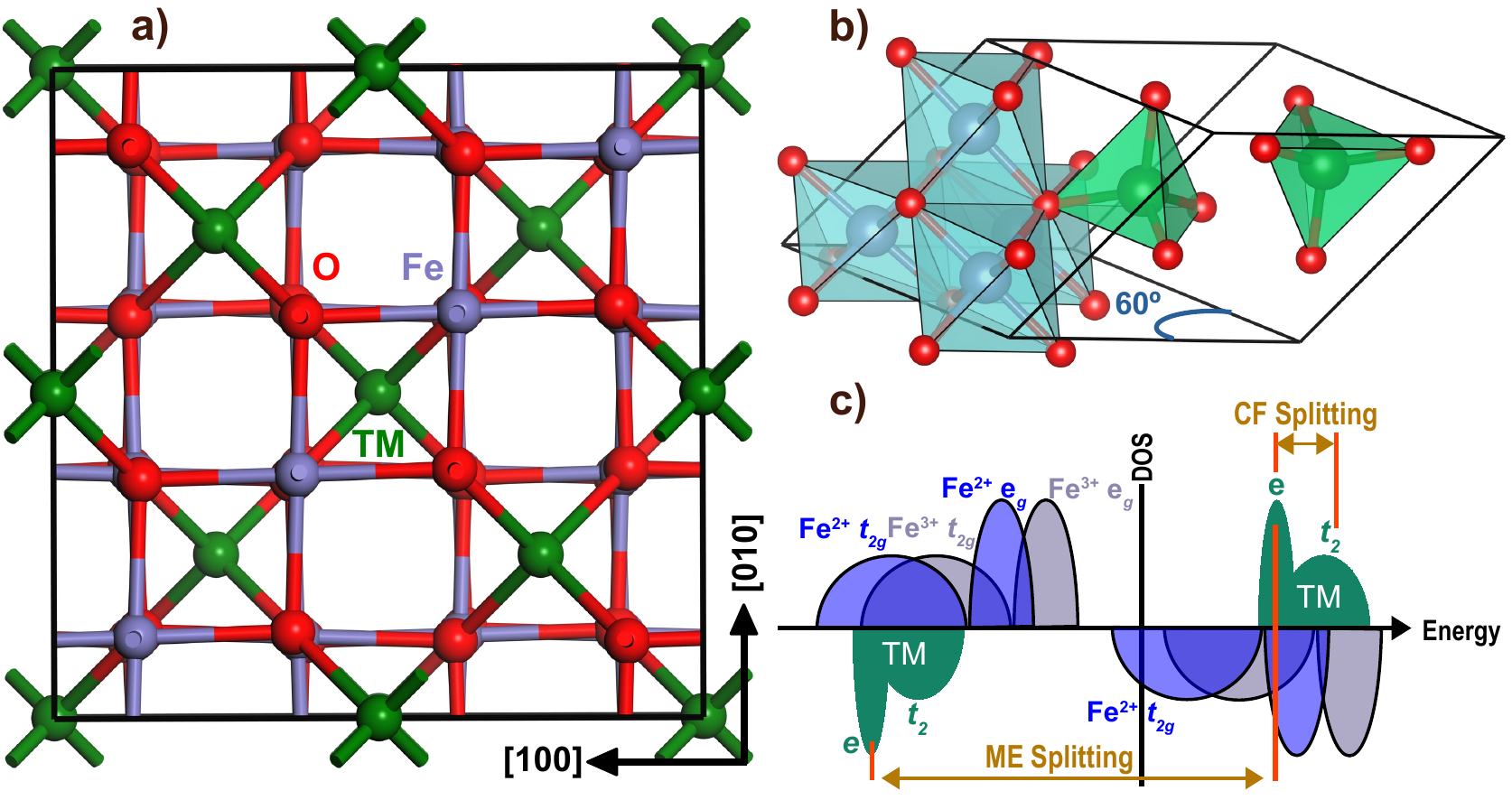}
            \caption{\label{figure:1}(a) The schematic presentation of the cubic (conventional) $Fd\bar3m$ cell of the \ch{TMFe2O4} compounds. (b) Primitive cell of the same structure in the polyhedron presentation. The TM ions are tetrahedrally coordinated while the Fe ions are octahedrally coordinated. (c) The schematic density of states of the representative \ch{Fe3O4} structure. Crystal field (CF) and the magnetic exchange (ME) are also demonstrated. For other compounds, the partial density of states of the octahedrally coordinated TM cations moves closer to the Fermi level based on occupancy.}
         \end{figure*}

We carried out spin-polarized density functional theory (DFT) calculations within the projector augmented wave formalism\cite{Kresse1999} as implemented in VASP code\cite{Kresse1996b, Kresse1996a} with an energy cut-off of $520$ \si{\eV} for geometry optimization. At this stage, we used a Brillouin zone sampling of a mesh generated by $9 \times 9 \times 9$ Monkhorst-Pack grid to relax the primitive cell (Figure \ref{figure:1}b) of the \ch{TMFe2O4} compounds to forces smaller than $0.01$ \si{\eV\per\angstrom}. We also applied a GGA$+U$ correction\cite{Dudarev1998, Liechtenstein1995} with an on-site Coulomb interaction term of $U = 3.5$ \si{\eV} and on-site exchange interaction of $J = 0.5$ \si{\eV} for all 3d TM ions, and $U = 3$ \si{\eV} and $J = 1$ \si{\eV} o for all 4d TM ions to improve the electronic description arising from the strong localization of d electrons throughout all calculations. These $U_{\mathrm{eff}}$ values reproduce the measured magnetic ordering and the electronic structure for \ch{Fe3O4}\cite{Fonin2007} and \ch{MoFe2O4}.\cite{Ramdani1985} The validity of these $U$ and $J$ values were examined and confirmed in \textbf{Figures S1}--\textbf{S3}. Furthermore, the use of a uniform $U$ and $J$ values for all 3d and 4d TM ions offers a straightforward comparison within the entire materials’ family.\cite{Gopal2006} The structural descriptions and the proof of the stability of the antiferromagnetic phase of the compounds studied here have been published elsewhere.\cite{Assadi2019} The elastic tensor was calculated using conventional $Fd\bar3m$ unitcell containing $56$ atoms based on the strain-stress method as implemented in VASP\cite{Wu2005} and extracted using MechElastic script.\cite{Singh2018} We employed central differences with a step size of $0.015$ \si{\angstrom}.

For calculating the density of states (DOS) and the transport properties, we used an ultra-fine $20 \times 20 \times 20$ Monkhorst-Pack grid in conjunction with an energy cut-off of $650$ \si{\eV}. This Monkhorst-Pack grid generated $7700$ unique irreducible k-points with a tight spacing of $\sim 0.01$ \si{\per\angstrom}. To ensure the ultimate accuracy, for smearing, we utilized the tetrahedron method with Blöchl correction. We then calculated the Seebeck coefficient using the BoltzTraP2 code,\cite{Madsen2018} which solves the linearized Boltzmann transport equation within the constant relaxation time ($\tau$) approximation, in which $\tau$ is assumed to be independent of temperature ($T$) and electron's energy ($E$). BoltzTrap2, therefore, only relies on the DFT calculated band and $k$-dependent quasiparticle eigenvalues as input. The assumption of a $T$ and $E$ independent $\tau$ results in a simple and tractable form of the equations for $S$, $\sigma$ and $\kappa_{\mathrm{e}}$.\cite{Scheidemantel2003} The constant relaxation time approximation, despite its simplicity, predicts $S$ values that match well with experiments, and is widely adopted in the high throughput theoretical search of novel thermoelectric materials.\cite{Chen2016, Gorai2017} The success of this approximation may stem from the fact that in doped semiconductors, such as the ones discussed here, carriers' relaxation time (and mobility) does vary very little with temperatures near and above ambient.\cite{Szmyd1990, Lovejoy1995, Singh2010} Constant relaxation time approximation, however, fails to describe the Seebeck coefficient for those materials in which the electron relaxation time is strongly energy-dependent such as Li.\cite{Xu2014} In Li the rapidly increasing DOS across the Fermi energy is the cause of the deviation from constant relaxation time approximation.

       \section{RESULTS AND DISCUSSION}
       \subsection{Search for Flat Band}
Based on the energy-independent carrier relaxation time and parabolic band approximations, the Seebeck coefficient of a TE material is approximately proportional to the density-of-states effective mass ($m_D^{\ast}$).\cite{Singh2000, Levin2016} Higher $m_D^{\ast}$ originate from sharp peaks in the density of state (DOS) near band edges which indicate flat and dispersionless bands.\cite{Mori2013, Yabuuchi2013} Therefore, in this section, we examine the DOS of the 3d TM and 4d TM containing \ch{TMFe2O4} compounds.

\begin{figure*}[!htb]
            \centering
            \includegraphics[width=0.9\columnwidth]{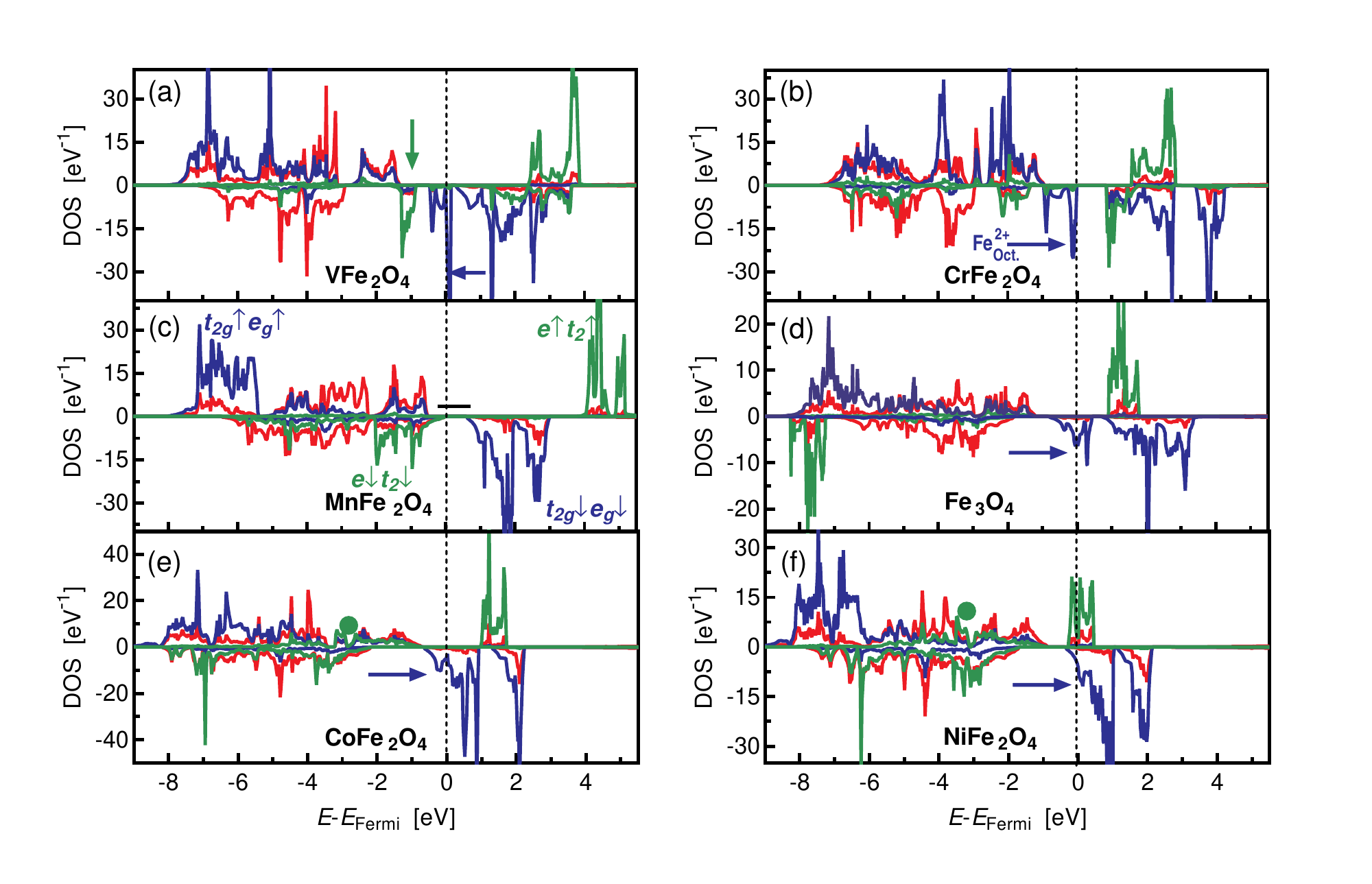}
            \caption{\label{figure:2}Partial density of states of 3d TM containing-\ch{TMFe2O4} compounds. Panels (a) through (f) correspond to \ch{VFe2O4} through \ch{NiFe2O4}, respectively, arranged by TM's atomic number. The blue, green, and red lines denote \ch{Fe} 3d, TM 3d, and \ch{O} 2p states, respectively.}
           \end{figure*}

Figure \ref{figure:2} shows the DOS of the 3d TM containing \ch{TMFe2O4} compounds. One general feature in all compounds is a sizeable magnetic exchange that separates the spin-up ($\uparrow$) channels from the spin-down ($\downarrow$) channels. Moreover, for the octahedrally coordinated Fe ions, the crystal field splits each spin channel into lower triply degenerate $t_{2g}$ and higher doubly degenerate $e_g$ states. For the tetrahedrally coordinated TM ions, this splitting is reversed to lower doubly degenerate $e$ and higher triply degenerate $t_2$ states. A schematic of the splittings is provided in Figure \ref{figure:1}c. As seen in Figure \ref{figure:2}, \ch{Fe} ions in all \ch{TMFe2O4} undergo a charge disproportionation into \ch{Fe^3+} and \ch{Fe^2+}, except in \ch{MnFe2O4}, which will be discussed shortly. For the high-spin \ch{Fe^3+}, the spin-up channel of the $t_{2g}$ and $e_g$ states (indicated with blue lines) are all filled and located at the bottom of the valence band ($\sim -8$ \si{\eV} $< E <$ $\sim -6$ \si{\eV}), while the spin-down channel is completely empty. For the high-spin \ch{Fe^2+} ions, one electron, however, occupies the spin-down $t_{2g}$ states, which are marked with blue arrows in Figure \ref{figure:2}. Given the dominance of this peak at the valence band top, its vicinity to the Fermi level, and its sharpness can determine $S$.

\ch{Fe} charge disproportionation in the majority of the 3d \ch{TMFe2O4} compounds dictates that all the tetrahedral TM ions are of +3 oxidation state except for \ch{Mn}. In \ch{MnFe2O4}, \ch{Mn} adopts the more stable \ch{Mn^2+} state leaving all \ch{Fe} in the +3 oxidation state. Consequently, one spin channel of each of the \ch{Mn} ($e^2\downarrow t_2^3\downarrow$) and \ch{Fe} ($t_{2g}^3\uparrow e_g^2\uparrow$) ions become fully occupied, while the respective opposite channels, which are separated by a magnetic exchange interaction, remain empty. Such an electronic configuration, which agrees rather well with earlier computational investigations,\cite{Huang2013} creates a gap of $0.59$ \si{\eV}, which is marked with a black bar in Figure \ref{figure:2}c.

Moreover, in all 3d \ch{TMFe2O4} compounds, the TM ions have their $e$ and the $t_2$ states progressively filled. As marked with a green arrow in Figure \ref{figure:2}a, for \ch{VFe2O4}, \ch{V^3+} has an electronic configuration of $e^2 t_2^0$ of which the $e^2$ electrons occupy the top of the valence band at $\sim -1.2$ \si{\eV} and oppose the spin direction of the filled \ch{Fe} states constituting a ferrimagnetic alignment. As the 3d TM ions move along the row and more electrons occupy the spin-down of $e$ and $t_2$ states, the occupied TM electrons move to lower energies. This trend is more evident in \ch{Fe3O4} (Figure \ref{figure:2}d) than in all other compounds for which the filled $e^2 t_2^3$ states of the tetrahedral \ch{Fe} are almost at the same energy level as the filled $t_{2g}^3 e_g^2$ states of the octahedral \ch{Fe^3+} at the bottom of the valence band ($\sim -7$ \si{\eV}). For \ch{CoFe2O4} and \ch{NiFe2O4}, the spin-down channel of \ch{Co} and \ch{Ni} remains at the bottom of the valence band while the spin-up channel gets progressively filled as marked with green circles in Figure \ref{figure:2}e and Figure \ref{figure:2}f.

\begin{figure*}[!htb]
           \centering
            \includegraphics[width=0.9\columnwidth]{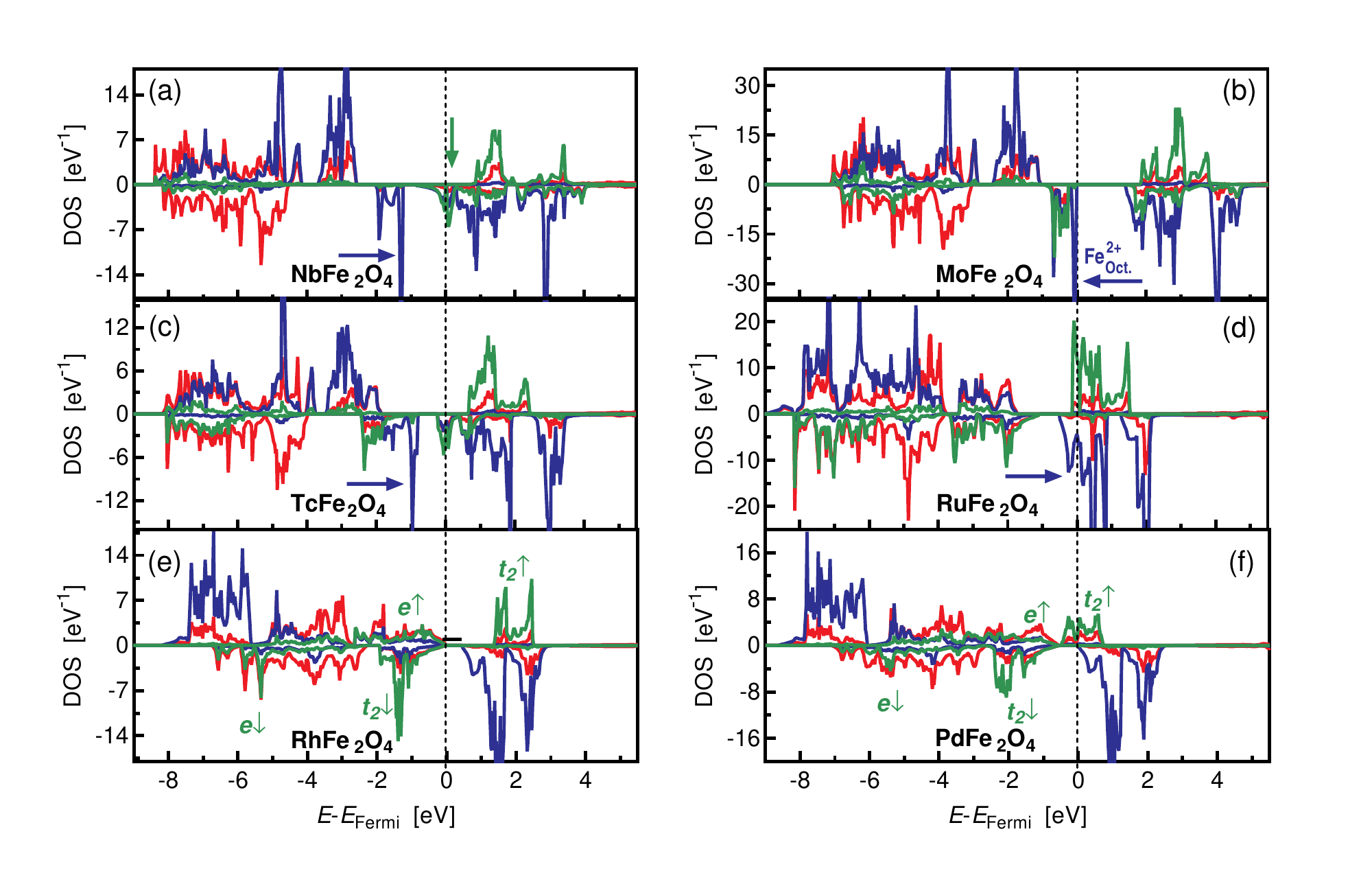}
            \caption{\label{figure:3} Partial density of states for the 4d TM-containing \ch{TMFe2O4} compounds. Panels (a) through (f) correspond to \ch{NbFe2O4} through \ch{PdFe2O4}, respectively, arranged by TM's atomic number. The blue, green, and red lines denote \ch{Fe} 3d, TM 4d, and \ch{O} 2p states, respectively.}
           \end{figure*}

The DOS for the 4d \ch{TMFe2O4} compounds, presented in Figure \ref{figure:3}, shows several similarities with those of the 3d counterparts; First, the \ch{Fe} ions are all in high-spin state and experience a sizeable magnetic exchange; Second, the Fermi level is dominated by the spin-down $t_{2g}$ states of the \ch{Fe^2+} ions (marked with blue arrows) except for the latter \ch{RhFe2O4} and \ch{PdFe2O4}; Third, the 4d TM ions adopt an antiferromagnetic alignment to the \ch{Fe} ions. The DOS of the 4d \ch{TMFe2O4} compounds, however, differ noticeably from those of the 3d containing \ch{TMFe2O4} in one aspect, and that is the 4d TM ions have smaller net magnetization than their 3d counterparts. The reduced magnetization can be attributed to the higher \ch{TM}--\ch{O} covalency in the case of the 4d elements.\cite{Assadi2019, Shirsath2019} Additionally, for \ch{RhFe2O4} and \ch{PdFe2O4}, \ch{Rh} and \ch{Pd} ions adopt the $+2$ oxidation state, leaving all \ch{Fe} ions as \ch{Fe^3+}. Consequently, there are no occupied spin-down \ch{Fe} $t_{2g}$ states in these compounds below the Fermi level. As shown in Figure \ref{figure:3}e, \ch{Rh^2+} adopts the $e^2\downarrow t_2^3\downarrow e^2\uparrow$ electronic configuration. The crystal field splitting between $e\uparrow$ and $t_2\uparrow$ creates a gap of $\sim 0.48$ \si{\eV}, which is marked with a black bar. \ch{Pd^2+}, as shown in Figure \ref{figure:3}f, adopts the $e^2\downarrow t_2^3\uparrow e^2\uparrow t_2^1\uparrow$ electronic configuration, and as a result, the Fermi level lies in the middle of the \ch{Pd} $t_2\uparrow$ states. Similar to the 3d TM containing compounds, the shape and position of the spin-down channel of the \ch{Fe^2+} ion may provide the necessary condition for high $S$ in 4d TM based \ch{TMFe2O4} compounds.

       \subsection{Thermoelectric Properties}
\begin{figure*}[!htb]
           \centering
            \includegraphics[width=0.9\columnwidth]{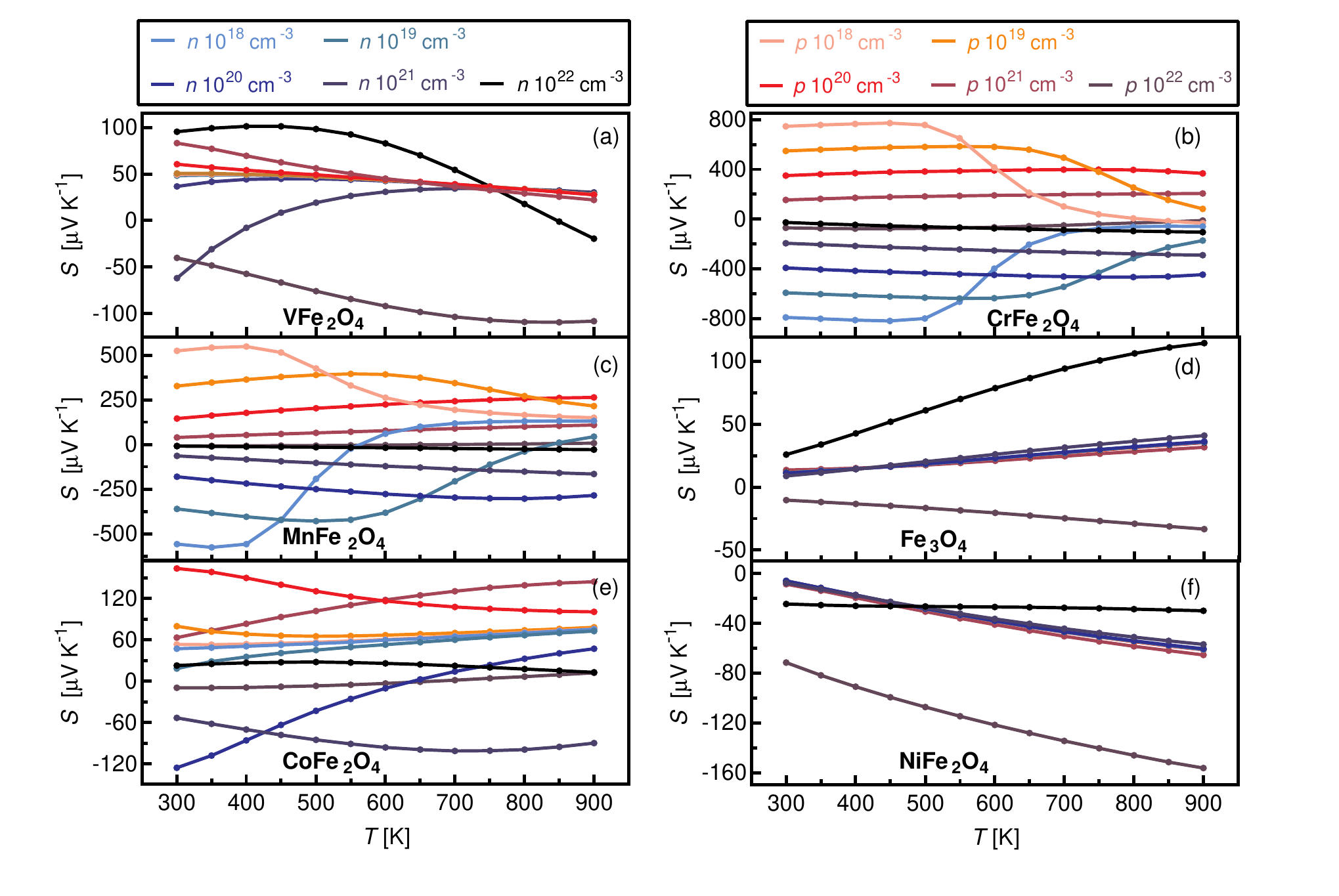}
            \caption{\label{figure:4}The predicted $S$ values for 3d-based \ch{TMFe2O4} compounds as a function of temperature and carrier doping. Panels (a) through (f) correspond to \ch{VFe2O4} through \ch{NiFe2O4}, respectively, arranged by TM's atomic number.}
            \end{figure*}

\begin{figure*}[!htb]
           \centering
            \includegraphics[width=0.9\columnwidth]{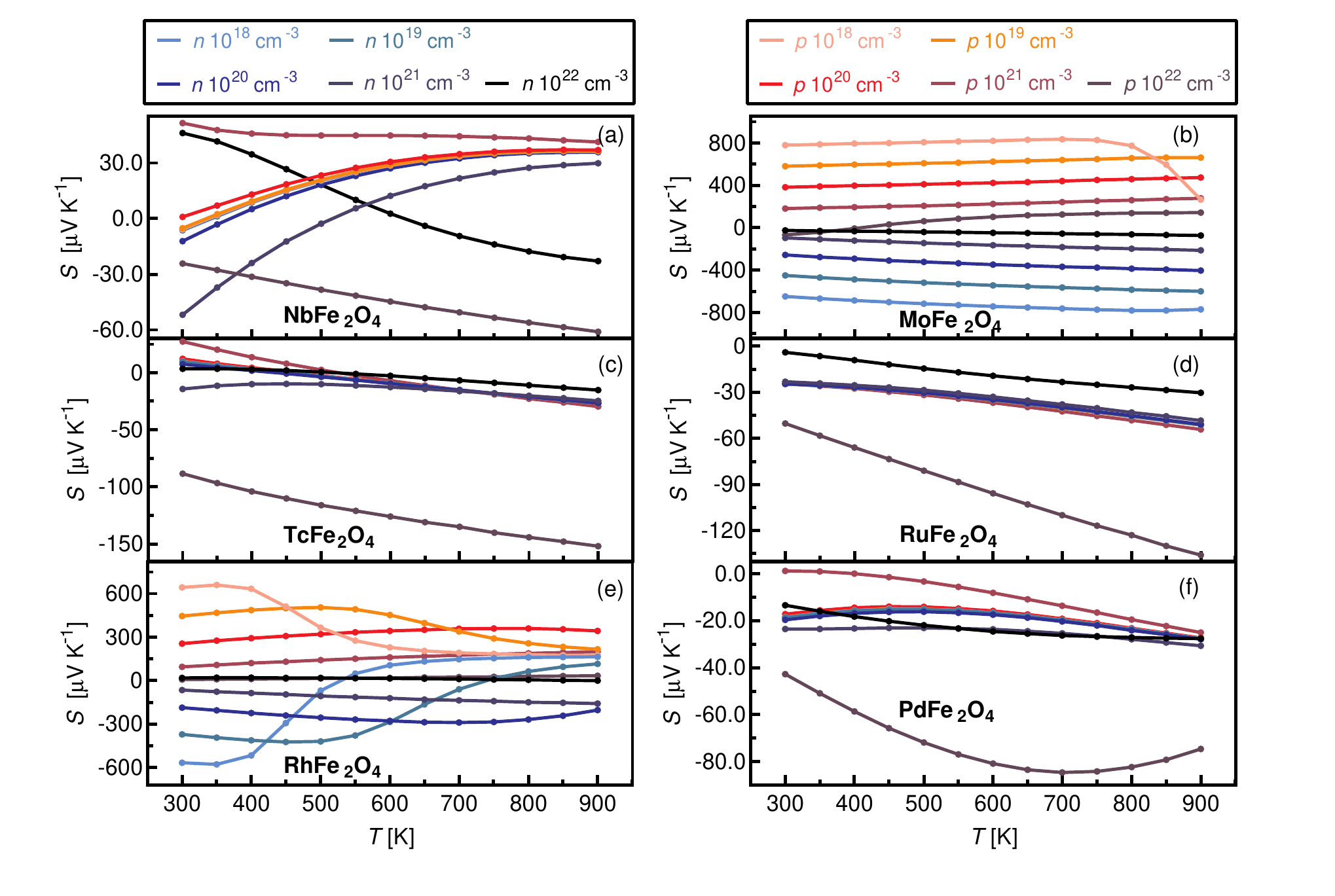}
            \caption{\label{figure:5}The predicted $S$ values for 4d-based \ch{TMFe2O4} compounds as a function of temperature and carrier doping. Panels (a) through (f) correspond to \ch{NbFe2O4} through \ch{PdFe2O4}, respectively, arranged by TM's atomic number.}
            \end{figure*}

Having the DOS calculated in the previous section, now we present the predicted Seebeck coefficients. The $S$ values as a function of the carrier doping and temperature for 3d and 4d containing \ch{TMFe2O4} compounds are presented in Figure \ref{figure:4} and Figure \ref{figure:5}, respectively. $S$ values as a function of the shift in the chemical potentials are shown in \textbf{Figures S4} and \textbf{S5}. For various doping levels, the $S$ value for \ch{VFe2O4}, \ch{Fe3O4}, \ch{CoFe2O4}, \ch{NiFe2O4}, \ch{NbFe2O4}, 
\ch{TcFe2O4}, \ch{RuFe2O4}, and \ch{PdFe2O4} falls approximately into the interval of $\pm 150$ \si{\micro\V\per\K}. The DOS indicates that these compounds are metallic or half-metallic, for which the effect of carrier doping is not as potent in influencing $S$. Among band insulators, \ch{MnFe2O4} and \ch{RhFe2O4} achieve higher $S$ values of $\pm 600$ \si{\micro\V\per\K} at $300$ \si{\K} when lightly doped with either n or p carriers at concentrations of $10^{18}$ and $10^{19}$ carriers $\mathbin{/}$ \si{\cm\cubed}. For \ch{MnFe2O4} and \ch{RhFe2O4}, $S$, nonetheless, falls rapidly with increasing doping level and temperature, especially for $T > 400$ \si{\K}. The best performing compounds are, however, the remaining band insulators, \ch{CrFe2O4}, and its 4d counterpart \ch{MoFe2O4}. For \ch{CrFe2O4}, at the low $10^{18}$ \si{\per\cm\cubed} n-type doping, $S$ is $-791$ \si{\micro\V\per\K} at $T = 300$ \si{\K} and reaches a minimum of $-819$ \si{\micro\V\per\K} at $T = 450$ \si{\K}. At the same level of p-type doping, $S$ is $746$ \si{\micro\V\per\K} at $T = 300$ \si{\K} and peaks to $772$ \si{\micro\V\per\K} at $T = 450$ \si{\K}. For \ch{MoFe2O4}, at $10^{18}$ \si{\per\cm\cubed} of n-type carrier doping, $S$ is $-646$ \si{\micro\V\per\K} at $T = 300$ \si{\K} with a minimum of $-779$ \si{\micro\V\per\K} at $T = 850$ \si{\K}. At the same level of p-type carrier doping, $S$ is $778$ \si{\micro\V\per\K} at $T = 300$ \si{\K} and peaks at $835$ \si{\micro\V\per\K} at $T = 700$ \si{K}. The $S$ values slightly fall by $\sim 50$ \si{\micro\V\per\K} for $10^{19}$ \si{\per\cm\cubed} carrier doping in \ch{CrFe2O4} and \ch{MoFe2O4}. The predicted $S$ values for these compounds are comfortably twice as large as that of \ch{Bi2Te3} ($\sim 250$--$260$ \si{\micro\V\per\K} at room temperature.)\cite{Jeon1991, Takashiri2007}

Furthermore, for the best performing \ch{CrFe2O4} and \ch{MoFe2O4}, $S$ falls rapidly with excessive carrier dopings beyond $10^{20}$ \si{\per\cm\cubed}. For instance, at $10^{22}$ \si{\per\cm\cubed} of both p and n-type doping, the absolute $S$ value falls to an average of $\sim 60$ \si{\micro\V\per\K} in \ch{CrFe2O4}, and $\sim 64$ \si{\micro\V\per\K} in \ch{MoFe2O4}. The substantial $S$ values in lightly doped \ch{CrFe2O4} and \ch{MoFe2O4} and their descent for more massive carrier doping stem from the sharp \ch{Fe^2+} 3d peak below the valence band maximum of both compounds---marked with blue arrows in Figure \ref{figure:2}b and Figure \ref{figure:3}b, respectively. Through light doping, the Fermi level can be ever slightly adjusted so that it falls close to the peak. For instance, in \ch{CrFe2O4}, $10^{18}$ \si{\per\cm\cubed} n-type doping shifts the Fermi level $-9.90 \times 10^{-4}$ \si{\eV} while $10^{22}$ \si{\per\cm\cubed} n-type doping shifts the Fermi level $-1.53$ \si{\eV} at $T = 300$ \si{\K}. We can, therefore, see that more massive doping moves the \ch{Fe} 3d peak too far away from the Fermi level and diminishes its favorable effect on $S$ (\textbf{Figure S6}). A similar trend can be seen for p-type doping and in \ch{MoFe2O4} (\textbf{Figure S7}).

       \subsection{Best Performing \ch{CrFe2O4} and \ch{MoFe2O4}}
\begin{figure*}[!htb]
           \centering
            \includegraphics[width=0.9\columnwidth]{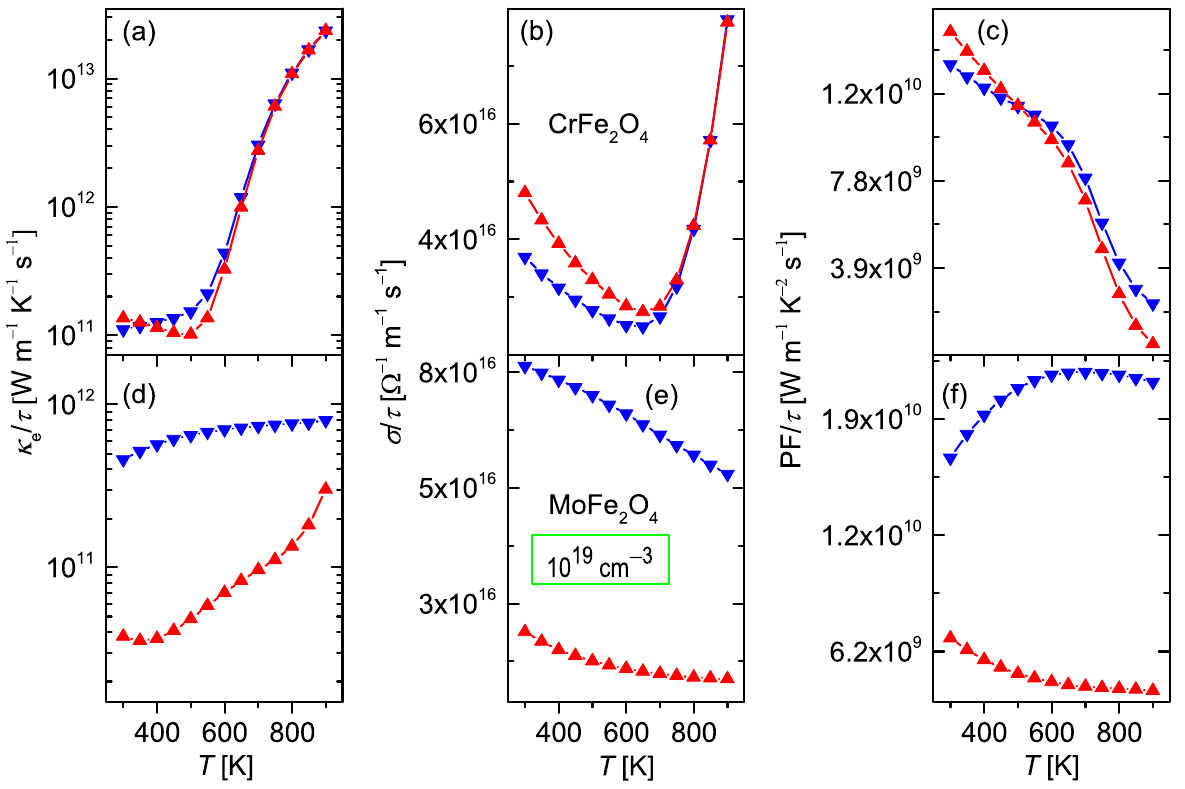}
            \caption{\label{figure:6}Upper row panels show the electronic contribution to the thermal conductivity per relaxation time ($\kappa_{\mathrm e}\mathbin{/}\tau$) (a), the electric conductivity per relaxation time ($\sigma\mathbin{/}\tau$) (b), and the power factor per relaxation time (PF$\mathbin{/}\tau$) (c) for \ch{CrFe2O4}. The lower row panels (d), (e), and (f) show the same quantities for \ch{MoFe2O4}. Red and blue symbols indicate hole and electron doping, respectively. The carrier concentration is $10^{19}$ carriers$\mathbin{/}$\si{\cm\cubed}.}
           \end{figure*}

Given that in the previous section, we predicted that the semiconducting \ch{CrFe2O4} and \ch{MoFe2O4} would have the highest Seebeck coefficient among the investigated compounds, their electrical conductivity, and the electronic contribution to the thermal conductivity are examined here. Accordingly, Figure \ref{figure:6} shows $\kappa_{\mathrm{e}}\mathbin{/}\tau$, $\sigma\mathbin{/}\tau$, and PF$\mathbin{/}\tau$, for the lightly doped, $10^{19}$ carriers$\mathbin{/}$\si{\cm\cubed}, \ch{CrFe2O4}, and \ch{MoFe2O4}. Here, $\tau$ is the mean relaxation time used in the BoltzTrap2 calculations. Moreover, we chose to show the results for lightly doped compounds as light doping is easier to achieve experimentally without running into dopant solubility problems. In \ch{CrFe2O4}, $\kappa_{\mathrm{e}}\mathbin{/}\tau$, $\sigma\mathbin{/}\tau$, PF$\mathbin{/}\tau$, for both light hole and light electron doping, have similar values and follow the same trend. $\kappa_{\mathrm{e}}\mathbin{/}\tau$ starts at $\sim 1.2 \times10^{11}$ \si{\W\per\m\per\K\per\s} at $T = 300$ \si{\K} and increases sharply by two orders of magnitude with the rising temperature at $T = 900$ \si{\K}. It is worthy of note that for a compound in which the Fermi level crosses the valence band, such as \ch{VFe2O4} of Figure \ref{figure:2}a, $\kappa_{\mathrm{e}}\mathbin{/}\tau$ is nonetheless two orders of magnitude higher than that of \ch{CrFe2O4} at the same doping level (\textbf{Figure S8}). $\sigma\mathbin{/}\tau$, after an initial dip, abruptly rises by twofold for $T > \sim 500$ \si{\K}, indicating a semiconducting behavior. Despite the increase in $\sigma\mathbin{/}\tau$ with temperature, PF$\mathbin{/}\tau$, nonetheless, drops by $\sim 1$ order of magnitude from its room temperature value with the rising temperature at $T = 900$ \si{\K}. This drop in PF$\mathbin{/}\tau$ is caused by the downward trend of $S$ with the temperature at $T > 600$ \si{\K} for \ch{CrFe2O4} doped at carrier concentrations of $10^{19}$ \si{\per\cm\cubed} (Figure \ref{figure:4}b). $S(T)$ is, nonetheless, nearly flat for higher doping levels in \ch{CrFe2O4}, which indicates this drop is milder for higher doping levels (\textbf{Figures S9} and \textbf{S10}). For \ch{MoFe2O4}, $\sigma\mathbin{/}\tau$, PF$\mathbin{/}\tau$ are generally a few times higher for n doping than for p doping. Furthermore, for both n and p doping, $\sigma\mathbin{/}\tau$ and PF$\mathbin{/}\tau$ vary moderately with temperature, indicating that a reasonably high power factor can be maintained even when the operating temperature varies.

To relate the values presented in Figure \ref{figure:6} to experimental measurements, we should estimate $\tau$. $\tau$ can be estimated from experimental carrier mobility ($\mu$) measurements via $\tau = \mu m^{\ast}\mathbin{/} q$, in which $q$ and $m^{\ast}$ are the carrier's charge and band effective mass, respectively. For the p-type \ch{MoFe2O4}, the hole relaxation time ($\tau_{\mathrm{h}}$) has already been measured to be $\sim 10^{-13}$ \si{\s} at room temperature for polycrystalline bulk samples.\cite{Ramdani1985, Gupta1979} Based on this $\tau_{\mathrm{h}}$ value, $\kappa_{\mathrm{e}}$ is estimated to be $3.78$ \si{\milli\W\per\K\per\m} at $T = 300$ \si{\K} and $30.14$ \si{\milli\W\per\K\per\m} at $T = 900$ \si{\K}. These values are rather small for typical oxides---for instance, $\kappa_{\mathrm{e}}$ is $\sim 1500$ \si{\milli\W\per\K\per\m} for doped \ch{SrTiO3} at ambient\cite{Wang2013}---indicating the minor role of electrons in heat transport. Furthermore, the room temperature $\sigma$ is estimated to be $20.37$ \si{\per\ohm\per\cm}, which is lower than that of most thermoelectric oxides (\textbf{TABLE S1}) and is in par with some other excellent thermoelectric materials such as \ch{ReSi_{1.75}}.\cite{Gottlieb1995, Inui2005} Higher conductivity can nonetheless be achieved by higher doping level (\textbf{Figures S9} and \textbf{S10}). The room temperature PF is estimated to be $689.81$ \si{\micro\W\per\K\squared\per\m}, which is higher than that of most oxides (\textbf{TABLE S1}). Assuming $\tau_{\mathrm{h}}$ remains constant with varying temperature, the power factor, at $T = 600$ \si{\K}, would be $455.67$ \si{\micro\W\per\K\squared\per\m}, which only shows a minor drop with respect to the PF value at ambient. Consequently, lightly doped p-type \ch{MoFe2O4} is anticipated to be an excellent choice for room and low temperature ($T < 600$ \si{\K}) applications.

Although Figure \ref{figure:6} indicates that $\sigma\mathbin{/}\tau$ and PF$\mathbin{/}\tau$ are $\sim 6$ times higher for n-doped \ch{MoFe2O4} than the p-doped compound, $\sigma$, and PF themselves may not be this high. That is because $\tau$ for electrons and holes is quite different and critically depends on the band effective mass. The band effective hole and electron masses, for \ch{MoFe2O4}, are $0.45 m_0$ and $2.52 m_0$, respectively, as calculated in \textbf{Figures S11}--\textbf{S13}, and \textbf{TABLE S2} ($m_0$ is electron mass at rest). Given that electrons are $\sim 5$ times heavier than holes, we should be somewhat conservative in predicting the thermoelectric performance of the n-type \ch{MoFe2O4}. The same forecast is also valid for \ch{CrFe2O4} for which the electron effective mass is $\sim 5$ times heavier than the hole effective mass.

Both \ch{CrFe2O4} and \ch{MoFe2O4} have been experimentally synthesized, and optically and electronically characterized, indicating their feasibility for thermoelectric applications.\cite{Chambers2017, Ramdani1985} Here, we further investigate the dynamic stability of these compounds. A material is dynamically stable if it passes the Born-Huang criteria.\cite{Born1955} These criteria state that the Gibbs free energy of any stable crystal is minimum compared to any other state induced by an infinitesimal strain. Fulfilling this requires that the $6 \times 6$ elastic stiffness matrix $C_{ij}$ to be positive definite, that is, all the eigenvalues of $C_{ij}$ are positive, while the $C_{ij}$ matrix itself is symmetric. Furthermore, for the cubic systems, such as the spinel structure, the following criteria must also be met: $C_{11}-C_{12}> 0$; $C_{11} + 2C_{12} > 0$; and $C_{44} > 0$.\cite{Mouhat2014} Table \ref{table:1} shows the unique non-zero stiffness matrix elements for \ch{CrFe2O4} and \ch{MoFe2O4}, along with the corresponding Debye temperatures ($\theta_{\mathrm{D}}$). Both of these compounds meet the Born stability criteria.
        
         \begin{table}
         \caption{The unique nonzero elements of the stiffness matrix along with the Debye temperature for \ch{CrFe2O4} and \ch{MoFe2O4.}}
         \label{table:1} 
         \begin{center}    
        \begin{tabular}{l c c}
        \hline
        \hline
           & \ch{CrFe2O4} & \ch{MoFe2O4} \\ [3pt]
       \hline
        $C_{11}$  (GPa) & $230.09$ & $175.48$ \\ [3pt]
        $C_{12}$  (GPa) & $135.94$ & $127.40$ \\ [3pt]
        $C_{44}$  (GPa) & $\phantom{0}69.87$ & $\phantom{0}59.20$ \\ [3pt]
        $\theta_{\mathrm{D}}$ (K) & $523.36$ & $405.23$ \\ [3pt]
       \hline
       \hline   
\end{tabular}
\end{center}
\end{table}
        
Finally, we would like to draw attention to the experimentally important fact that all investigated compounds here were of spinel structure in which all non-iron TM ions (except for \ch{Fe3O4}) were at the tetrahedral site (A site). Often, the site preference of the non-iron cation in ferrites depends on the synthesis method. In extreme cases where all the non-iron cations are located at the octahedral site (B site), the structure is referred to as an inverse spinel. In reality, any given spinel ferrite may be in an in-between case characterized by an inversion parameter. For instance, $\sim 60$ \si{\nm} thick \ch{MoFe2O4} deposited on \ch{MgAl2O4} [100] by pulsed laser deposition has an inverse spinel structure.\cite{Katayama2018} \ch{Mo}'s site preference in polycrystalline bulk samples could, nonetheless, be tuned by the sintering temperature.\cite{Ramdani1985} The higher the sintering temperature was, the more likely \ch{Mo} occupied the tetrahedral site. Generally, the site preference of the different cations in spinels can be fine-tuned by adjusting the strain (lattice mismatch)\cite{Fritsch2011}, self-doping\cite{Bahlawane2009} and annealing\cite{Ndione2014} in thin films; and selecting suitable precursors,\cite{Petrov1988} sintering temperature\cite{Ramdani1985} and nano-structuring\cite{Song2012} in bulk samples. The wealth of the experimental know-how in synthesizing ferrites can undoubtedly come handy in developing thermoelectric \ch{TMFe2O4}, especially for nanostructuring as a mean of reducing the lattice thermal conductivity and enhancing thermoelectric response.\cite{Azough2019}
       
       \section{CONCLUSIONS}
Using density functional band structure calculations and linearized Boltzmann transport equation, we surveyed the thermoelectric properties of twelve ferrite compounds of \ch{TMFe2O4} composition with spinel structure in which TM was either a 3d or 4d transition metal cations. We demonstrated that the absolute value of the Seebeck coefficient, at ambient conditions, can exceed $\pm600$ \si{\micro\V\per\K} in \ch{CrFe2O4},\ch{ MnFe2O4}, \ch{MoFe2O4}, and \ch{RhFe2O4} when lightly doped with electrons and holes at concentrations smaller than $10^{20}$ carriers$\mathbin{/}$\si{\cm\cubed}. Additionally, in these compounds, $S$ is the highest at room temperature and tapers off very moderately with rising temperatures up to $600$ \si{\K}. This behavior is starkly different from that of most thermoelectric oxides for which $S$ is minuscule at ambient and only becomes significant at temperatures higher than $800$ \si{\K}. Consequently, for p-type \ch{MoFe2O4} the thermoelectric power factor can reach $689.81$ \si{\micro\W\per\K\squared\per\m} at $300$ \si{\K}, and $455.67$ \si{\micro\W\per\K\squared\per\m} at $600$ \si{\K}. The unusually high $S$ for \ch{CrFe2O4} and \ch{MoFe2O4} is caused by the \ch{Fe^2+}'s sharp density of states peak in the minority spin channel just below the valence band maximum. The analysis performed here, by contributing to the understanding of thermoelectrics properties of oxides, will facilitate more extensive use of this class of materials for applications close to room temperature.       
              
       \section{CONFLICTS OF INTEREST}
The authors declare that there is no conflict of interest.
       
      \section{ACKNOWLEDGMENTS}
 Computational resources were provided by National Computational Infrastructure, Australia. J.J.G.M acknowledges the financial support from the China Postdoctoral Science Foundation under Grant No. 2018M643152.              
        
       \section{SUPPORTING INFORMATION}
Validity tests for the applied $U$ and $J$ values, $S$ vs. $\mu$ plots at different temperatures and as a function of the shift in the chemical potential, thermoelectric transport for \ch{VFe2O4} at $10^{19}$ carriers per \si{\cm\cubed} doping, \ch{CeFe2O4} and \ch{MoFe2O4} at $10^{20}$ and $10^{21}$ carriers per \si{\cm\cubed} doping, comparison of the thermoelectric performance with other oxides, the procedure for calculating the effective masses.

  \bibliography{ms}{}

\end{document}